\documentclass[journal]{IEEEtran}
\usepackage{graphicx,times}
\usepackage{subfigure}
\usepackage{array}
\usepackage{multirow}
\usepackage{float}
\usepackage{amsmath}

\usepackage[colorlinks,linkcolor=blue]{hyperref}

\makeatletter
\newcommand{\thickhline}{%
    \noalign {\ifnum 0=`}\fi \hrule height 1pt
    \futurelet \reserved@a \@xhline
}
\newcolumntype{"}{@{\hskip\tabcolsep\vrule width 1pt\hskip\tabcolsep}}
\makeatother

\ifCLASSINFOpdf
\else
\fi
\begin{document}
%
\title{Implicit Neural Representation Learning for \\ Hyperspectral Image Super-Resolution}
%
%
%


\author{Kaiwei~Zhang,~\IEEEmembership{Member,~IEEE,}
\thanks{Kaiwei Zhang is with the Institute of Image Communication and Network Engineering, Shanghai Jiao Tong University, Shanghai 200240, China (e-mail:zhangkaiwei@sjtu.edu.cn).}}

\maketitle

\begin{abstract}
Hyperspectral image (HSI) super-resolution without additional auxiliary image remains a constant challenge due to its high-dimensional spectral patterns, where learning an effective spatial and spectral representation is a fundamental issue. Recently, Implicit Neural Representations (INRs) are making strides as a novel and effective representation, especially in the reconstruction task. Therefore, in this work, we propose a novel HSI reconstruction model based on INR which represents HSI by a continuous function mapping a spatial coordinate to its corresponding spectral radiance values. In particular, as a specific implementation of INR, the parameters of parametric model are predicted by a hypernetwork that operates on feature extraction using convolution network. It makes the continuous functions map the spatial coordinates to pixel values in a content-aware manner. Moreover, periodic spatial encoding are deeply integrated with the reconstruction procedure, which makes our model capable of recovering more high frequency details. To verify the efficacy of our model, we conduct experiments on three HSI datasets (CAVE, NUS, and NTIRE2018). Experimental results show that the proposed model can achieve competitive reconstruction performance in comparison with the state-of-the-art methods. In addition, we provide an ablation study on the effect of individual components of our model. We hope this paper could server as a potent reference for future research.
\end{abstract}

\begin{IEEEkeywords}
Hyperspectral super-resolution, implicit neural representation, continuous functional mapping, hypernetwork.
\end{IEEEkeywords}
\IEEEpeerreviewmaketitle

\section{Introduction}
Hyperspectral imaging aims at capturing images of the same scene in many spectral bands over a wavelength range. The resulting hyperspectral image can reflect more intrinsic properties of object materials compared with commonly used multispectral image (MSI) such as RGB image. With the availability, it is widely applied in remote sensing \cite{goetz1985imaging,bioucas2013hyperspectral,melgani2004classification}, medical imaging \cite{lu2014medical}, and computer vision tasks \cite{pan2003face,xiong2020material,tarabalka2010segmentation,li2011hyperspectral,zhang2021s3net}. 
However, due to hardware and budget constraints of spectrometer sensors, HSI generally has lower resolution in spatial domain than RGB image. Additionally, the sensors acquire more exposure time, during which all other factors must be kept constant to ensure an acceptable signal-to-noise-ratio. These limitations impede the practical applications for HSI. Therefore, how to obtain an appropriate hyperspectral image with high resolution is a desirable study problem.

Super-resolution has been one of the most promising approaches among image reconstruction technologies, which is economical and efficacious without the hardware modification to recover more details from observed low-resolution images. According to the number of input images, HSI super-resolution methods can be divided into two categories: fusion based super-resolution \cite{xue2021spatial,zhang2020unsupervised,hu2021hyperspectral} (aka pansharpening) and single image super-resolution. The former merges a multispectral image with high spatial resolution and a HSI with low spatial resolution to generate a HSI with high spatial resolution, while the latter aims to reconstruct a HSI from its corresponding RGB image. Under experimental conditions, fusion based super-resolution methods can generally achieve superior performance than single image super-resolution. However, in practice, the low-resolution HSI and high-resolution multispectral image are captured by different CCD cameras and thus the consequent minute changes in imaging factors (e.g., optical filter, zoom lens and sensor shaking) would make it challenging to obtain two well-registered images. Hence, we will focus on the single image super-resolution which is proposed to overcome the unavailability of co-registered images by improving the spectral resolution of the multispectral image without additional auxiliary image.

\begin{figure}[!t]
  \centering
  \includegraphics[scale=0.22]{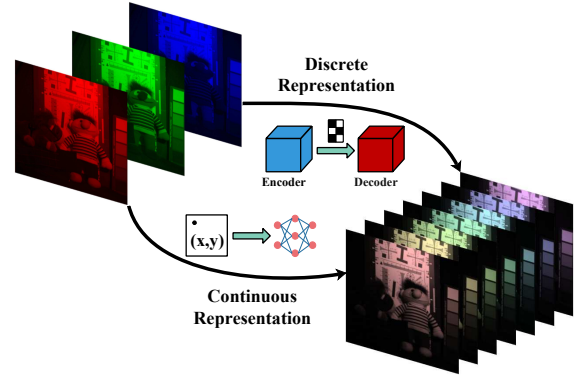}
  \caption{The hyperspectral image super-resolution process with discrete representation and continuous representation. Top branch: the discrete representation is an autoencoder structure in which the encoder maps the input into the code and the decoder maps the code to a reconstruction of high resolution. Bottom branch: the continuous representation is a continuous function that maps spatial coordinates $(x,y)$ to pixel values.}
\label{FIG_intro}
\end{figure}

Since the single image super-resolution is an ill-posed inverse problem, traditional methods generally employ different prior knowledge as regularization terms. Parmar \textit{et al.} \cite{parmar2008spatio} first propose a spectral reconstruction method in a sparse representation. In 2016, Arad and Ben-Shahar \cite{arad2016sparse} create a sparse dictionary of hyperspectral signatures and corresponding RGB projections for the first time by introducing hyperspectral prior. Zhang \textit{et al.} \cite{zhang2019computational} reconstruct the HSIs on the basis of the low-rank tensor recovery to boost the reconstruction performance. However, the hand-crafted priors are not adequate to reflect various characters of natural hyperspectral data. Recently, Convolutional Neaural Networks (CNNs) have infiltrated computer vision and enhanced dramatically the performance of various tasks, such as spatial super-resolution \cite{dong2015image,ledig2017photo}. Inspired by these researches, many CNN networks are also  designed to the task of single image spectral super-resolution. In \cite{mei2017hyperspectral}, Mei \textit{et al.} propose a 3D full CNN to extract spectrum features by exploiting the context of neighboring pixels and spectral bands. In \cite{jiang2020learning}, a spatial-spectral prior network is introduced to fully explore the correlation between spatial and spectral information of the HSI. He \textit{et al.} \cite{he2021spectral} utilize the spectral response function to divide the spectral similar bands into the same group to enhance the capability of spectral details extraction in CNN.

As illustrated in the top branch of Fig. \ref{FIG_intro}, deep learning based methods aim at learning a RGB-to-HSI discrete mapping of 2D arrays of pixels, so this mapping only reflect a sparse sampling of the spectral scene. However, our visual world is continuous and the spectral response function is actually a smooth curve. Thus, a natural idea comes: how to represent a hyperspectral image as a continuous function. Some related progress in implicit neural representation for 3D shape reconstruction \cite{sitzmann2019scene,mildenhall2020nerf}, image translation \cite{shaham2021spatially} and image up-sampling \cite{chen2021learning} provides new enlightenment for our work.

To this end, we propose a hyperspectral image representation model which reconstructs HSI by a continuous function mapping each spatial coordinate to a corresponding spectral value at that coordinate. As seen in the bottom branch of Fig. \ref{FIG_intro}, the continuous representation function is parameterized as a Multi-Layer Perceptron (MLP) with ReLU \cite{nair2010rectified}. In addition, we construct a hypernetwork which takes an image and returns weights and biases for the MLP. In this way, different input images yield different sets of parameters, and it in turn means the MLP is adaptive to different input image. With this diversity, the model becomes fully expressive. Specifically, we first encode spatial coordinates using sinusoids at different frequencies. Then, the spatial coding and input image are merged together to feed the MLP. In summary, the main contributions of our work are as follows:

$\bullet$ We propose a novel HSI reconstruction model based on INR for mapping spatial coordinates to spectral radiance intensities. The proposed model builds a bridge between discrete pixels and continuous representation in spectral domain.

$\bullet$ The continuous function is approximated by a MLP, whose parameters are predicted by a hypernetwork. Besides, periodic spatial encoding projects the pixel coordinates into a higher dimensional space for recovering more high frequency details.

$\bullet$ Experiments on CAVE, NUS, and NTIRE2018 datasets can verify the superiority of our model in comparison with the state-of-the-art methods.

\section{Related Work}
\subsection{Single Image Super-Resolution}
Without the help of auxiliary HSI of low-resolution, single image SR is a more challenging tasks than fusion based SR. In \cite{akgun2005super}, Akgun \textit{et al.} introduced the HSI acquisition model and projection onto convex sets (POCS) method \cite{bauschke1996projection} to HSI super-resolution task, which can boost the high-resolution reconstruction. Some methods usually take advantage of various prior knowledge or assumptions to restrain the super-resolution problem. Nguyen \textit{et al.} \cite{nguyen2014training} considered the spectral response functions of the camera as known conditions, and proposed a radial basis function network to model the reconstruction process from RGB images to their spectral imageries. Huang \textit{et al.} \cite{huang2014super} used the low-rank and group-sparse as the prior to design a novel method for absolving image blur phenomena. Moreover, other conventional approaches \cite{wang2017hyperspectral,li2016hyperspectral} focused on exploring sparse representation and dictionary learning. However, these methods usually have to solve the constrained optimization problem which is high computational complexity and time consuming. Besides, the hand-crafted priors are based on specific types of image characteristics without any consideration of real-world situations.

To overcome these drawbacks, deep learning based techniques have been introduced into the single image SR task recently. In the beginning, Xiong \textit{et al.} \cite{xiong2017hscnn} proposed to increase the resolution of RGB images in spectral domain and employed an end-to-end network to learn the residual between the generated result and ground truth HSI. Li \textit{et al.} \cite{li2018single} embedded the group-wise convolution and recursive structure into a grouped deep recursive residual network, thereby achieving high performance. Fu \textit{et al.} \cite{fu2020joint} connected a spectral network and a spatial network in series to learn the spectral nonlinear mapping and spatial details respectively. In \cite{arun2020cnn}, a 3D-CNN based encoding-decoding architecture was introduced to model the spatial-spectral prior for improving spectral as well as spatial fidelity of reconstructions. Hang \textit{et al.} \cite{hang2021spectral} proposed a spectral super-resolution network, including a decomposition sub-network based on spectral correlation and a self-supervised sub-network to improve the reconstruction result. For further exploring the correlation among spectral bands, Zheng \textit{et al.} \cite{zheng2021spectral} proposed a neighboring spectral attention module and embedded the spatial–spectral residual blocks into their networks.
However, these deep learning methods either lack model interpretation, or are treated as a data-driven black box that provides implementation for nonlinear discrete mapping. Instead, our method equips continuous function mapping with the implicit neural representation, achieving better performance.

\subsection{Implicit Neural Representation}

After a series of research, traditional discrete representations of 3D object shapes, surfaces, and scene structure have been replaced by continuous functions parameterized by MLPs. Compositional Pattern Producing Networks \cite{stanley2007compositional} can be seen as one of the earliest attempts of continuous functional representations that independently map pixel coordinates to signals in a specific domain.
Scene Representation Networks \cite{sitzmann2019scene} represented the continuous scene as an opaque surface, implicitly defined by a MLP that outputs a feature vector and RGB color at each world coordinate. Meanwhile, image formation was formulated as a differentiable ray-marching algorithm to decide where the surface is located.
In NeRF \cite{mildenhall2020nerf}, Mildenhall \textit{et al.} expanded the input 3D coordinate to continuous 5D scene representation with a 2D viewing direction, and the 5D neural radiance fields produced better renderings of high-frequency scene content than other voxel methods.
Sitzmann \textit{et al.} \cite{sitzmann2020implicit} proposed a implicit neural representation network with periodic activation functions and it was demonstrated to be suitable for representing complex signals.
In \cite{zhang2021holistic}, an image-based local structured implicit network was proposed to reconstruct accurate 3D object geometry, and a graph-based convolutional network (GCN) was designed to learn better scene context.

Instead of learning implicit neural representations for 3D geometry shape, recent works have demonstrated the potential of implicit neural representations for 2D images \cite{sitzmann2020implicit, zhang2021nerd} and 1D audios \cite{sitzmann2020implicit}.
For example, Local Implicit Image Function (LIIF) \cite{chen2021learning} predicted a pixel of the image by querying the nearest latent code which is a local ensemble of neighboring feature vectors.
The pixel-based representation ensures the smooth transition of each region in reconstructed image.
Inspired by LIIF, Tang \textit{et al.} \cite{tang2021joint} proposed a novel Joint Implicit Image Function (JIIF) representation for guided depth super-resolution to learn both the interpolation weights and values.
According to the UltraSR \cite{xu2021ultrasr}, it might lack evidence that a simple concatenation on coordinates and spatial encoding improve the quality of the output images, so deep coordinate fusion of spatial coordinates and periodic encoding are fed into implicit image function to utilize the high-frequency hints in arbitrary-scale SR. Different from previous methods that treat feature vector from the low-resolution image domain as an input of MLP, we employ a hypernetwork to parameterize the MLP by the feature map of LR image. The resulting spatially-varying parameters improve the generalization of the super-resolution task.

\section{Methodology}
\begin{figure*}[!htbp]
  \centering
  \includegraphics[scale=0.45]{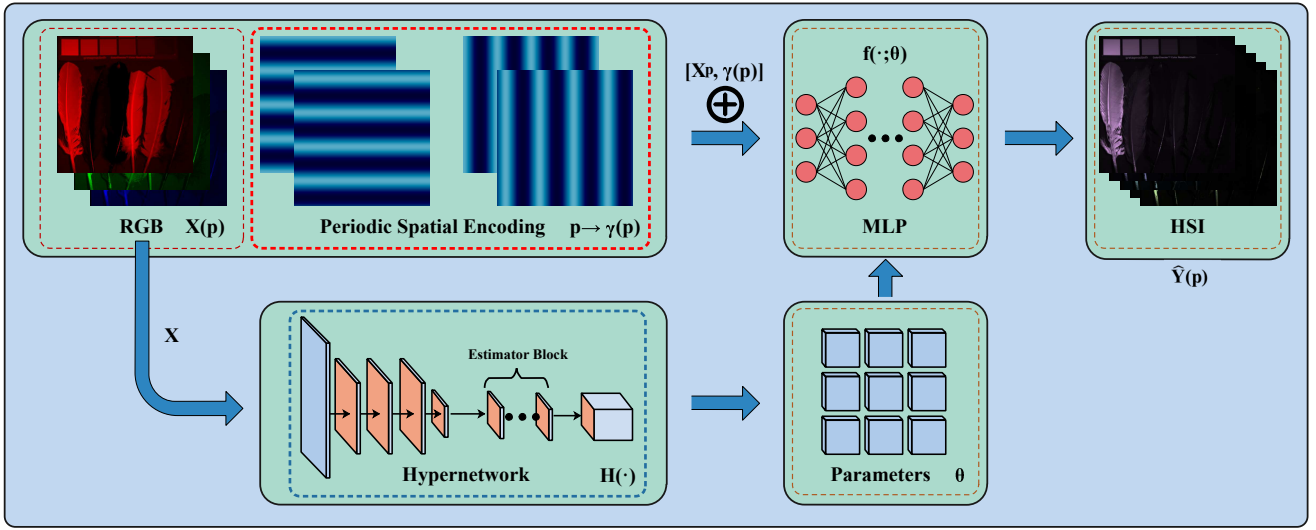}
  \caption{Architecture overview. Our model first processes the input at hypernetwork to produce a tensor of weights and biases $\theta$ for MLP. Then the MLP computes the final output from the input $[X_p, \gamma(p)]$.}
\label{FIG_Model}
\end{figure*}

In this work, we aim at combining hypernetworks and coordinate-based MLP to simulate implicit image function for HSI reconstruction, as illustrated in Fig. \ref{FIG_Model}.
The hypernetwork takes RGB image $X$ as input and produce parameters $\theta$ for an MLP $f(\cdot;\theta)$. 
Besides, we adopt deep Fourier features to encode the pixel coordinates as periodic spatial encoding. In practice, this encoding is a sinusoidal activation that maps a coordinate $p$ into a vector $\gamma(p)$. Note that the frequencies of the Fourier features is not kept fixed but covers in a multiplier range.
Finally, we concatenate the periodic spatial encoding with RGB image and feed them to the MLP.
Such fusion ensures that all hidden units have access to the target’s neighboring pixels and can fully explore the high-frequency cues. The resulting $\hat{Y}$ is the reconstructed HSI. Next, these will be introduced in detail.

\subsection{Model Formulation}
Let $X\in{R}^{W\times H\times 3}$ denote a RGB image, where $W$ and $H$ represent the width and height of its spatial size, respectively, and 3 is the number of color channels in RGB images. $Y\in{R}^{W\times H\times L}$ represents the ground truth HSI corresponding to $X$, Where $L>>3$ is the number of spectral bands in hyperspectral images. Since HSI super-resolution only focuses on improving the spectral resolution, the mapping function $M: R^3 \rightarrow R^L$ contains the same spatial size to reconstruct $\hat{Y}$ from $X$. In the training stage, the mapping is learned from the paired RGB and HSI in the training set, and it can be estimated by minimizing the following objective function:
\begin{equation}\label{eq1}
L1 = \left\| {Y-M(X)} \right\|_1,
\end{equation}
where $\left\|\cdot\right\|_1$ is the L1 norm. Current deep learning methods usually adopt L1 to impose the similarity between the upsampled output and the observed image.

From the imaging perspective, spectral value $Y(p)$ at the spatial position $p$ is divided by spectral bands into different wavelength interval $\lambda_1,...,\lambda_L$:
\begin{equation}\label{eq2}
Y(p)=\int _{\lambda_1,...,\lambda_L} R(p, \lambda )d\lambda,
\end{equation}
where $R(p, \lambda)$ denotes the spectral radiance at spatial position $p$ on wavelength $\lambda$. For RGB cameras, the sensors obtains the accumulation of different response in the spectral range. Thus, the RGB value $X_c(p)$ on different color channel can be expressed as:
\begin{equation}\label{eq3}
X_c(p)=\int _\Lambda R(p, \lambda )\Phi _c(\lambda )d\lambda,
\end{equation}
where $c\in \{R,G,B\}$ is the color channel, $\Lambda$ represents the spectral wavelength range, and $\Phi _c(\lambda )$ denotes the spectral response function at $c$ channel. Actually, in practical applications, the wavelength range is discretely sampled as wavelength intervals. Then RGB value $X_c(p)$ approximates to a discrete form:
\begin{equation}\label{eq4}
X_c(p)=\sum_{n=1}^{L} R(p, \lambda _n)\Phi _c(\lambda _n),
\end{equation}
where n denotes the index of spectral band. Since $c$ is the color band of images, $X_p$ can be the vector of $[R,G,B]$. Our method employ implicit neural representation to take as input the pixel’s coordinates $p=(x,y)$ and its color value $X_p$ and map them to signals:
\begin{equation}\label{eq5}
f(X_p, p; \theta ) = \hat{Y}(p),
\end{equation}
where $f(\cdot;\theta)$ is the learned mapping function by a MLP with ReLU activation layers, $\theta$ denote the weights and biases for the MLP, and $\hat{Y}(p)\in [0,1]^L$ is the reconstructed HSI. Specifically, the MLP is parameterized by a hypernetwork: 
\begin{equation}\label{eq6}
\theta =H (X),
\end{equation}
where $H(\cdot)$ denotes the hypernetwork. For general positional coordinate, $f:(x, y, X_p) \rightarrow \hat{Y}(p)$ is responsible for approximating a HSI at every real-valued coordinate $(x, y)\in[0, 1]^2$.

\subsection{Periodic Spatial Encoding}
\label{section_encod}
The key on the super-resolution task is to learn a good representation of the high frequency details. In the early application, LIIF directly feeds the coordinates and neighboring feature vectors into the implicit image function represented by a subsequent learnable MLP. However, in 3D scenes task, recent works \cite{mildenhall2020nerf,rebain2021derf,wizadwongsa2021nex,wu2021irem} have demonstrated that designed spatial encoding significantly improves performance of recovering fine-grained details. 
BARF \cite{lin2021barf} introduces a training strategy to only allow low frequency encodings to pass at the start of training and gradually activate the positional encodings of higher frequency bands during training. These researches proved that the encoding of pixel position can help focus to learning a high-fidelity representation.

Therefore, in this section, we perform periodic spatial encoding to map the positional coordinates to a higher dimensional space $R^{4N}(4N >2)$ before passing them to the fully-connected network. We let $\gamma_k (p)$ denote periodic spatial encoding from the space $R^2$ to $R^4$ at k-th frequency:
\begin{equation}\label{eq7}
\gamma_k (p)=[\cos(2^k\pi x), \sin (2^k\pi x), \cos(2^k\pi y), \sin (2^k\pi y)].
\end{equation}

Afterwards, we set $k=0,\dots,N-1$ to encode each k-th frequency component of the 2D pixel position $p = (x, y)$ as a vector of sinusoids with different frequencies:
\begin{equation}\label{eq8}
\gamma (p)=[\gamma_0 (p), \dots, \gamma_{N-1} (p)].
\end{equation}

At last, the sinusoidal encodings of spatial coordinates are concatenated with RGB color channels to augment the input of our subsequent MLP. Therefore, the input spatial encoding $\gamma (p)$ together with their corresponding pixel values are fed into the implicit image function shown as:
\begin{equation}\label{eq9}
f(X_p, \gamma (p); \theta ) = \hat{Y}(p).
\end{equation}

For this step, the $R^2$ linear spatial position is expanded into a augmented $R^{4N+3}$ input, which enables our subsequent MLP to more easily approximate to a higher frequency function.

\subsection{Parametric Model of Implicit Neural Representation}
\label{section_grid}
As illustrated in the bottom branch of Fig. \ref{FIG_Model}, this component of our architecture is the hypernetwork module for parameter generation. This RGB image will be processed by the strided-convolution layers first to reduce the spatial dimensions. Then the feature maps will be routed to the estimator block to estimate the MLP parameters. The estimator block consists of a convolutional group (without scaling transformation) and spatially-adaptive normalization layers, which ensures to generate a feature map with the same size as the previous feed. Meanwhile, it can extract deep features and probe abilities to characterize the image further. At the end of the hypernetwork, the final output of the last layer is a tensor with corresponding channel length to the weights and biases of the MLP.

\begin{figure}[!b]
  \centering
  \includegraphics[scale=0.7]{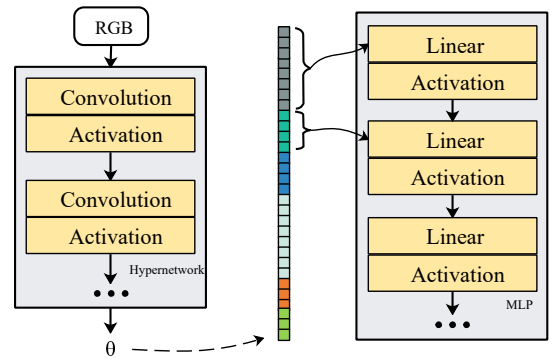}
  \caption{Parameters transfer. The hypernetwork is a convolutional network with ReLU activations that takes RGB image as input and produces the MLP parameters $\theta$. Then the $\theta$ is transfered to each layer of the MLP that is a fully-connected network with non-linearities activation layers.}
\label{FIG_PARA}
\end{figure}

Since spatially variable parameters are predicted by the convolutional estimator block, this makes the MLP adaptive to the input image, or in other words, the represent function depends on the input itself. Afterwards, the weights and biases are split up into different groups and transfered to each layer of the MLP in Fig. \ref{FIG_PARA}. The MLP consists of five hidden layers with Leaky-ReLU activation functions. In summary, the model parameterization method enables realistic image output with consistency edge and detailed structures.

\begin{figure}[t]
  \centering
  \includegraphics[scale=0.45]{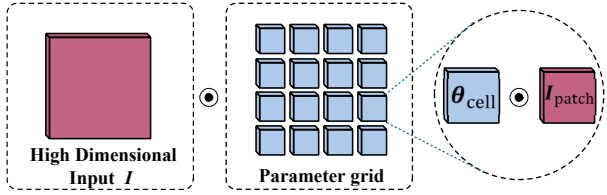}
  \caption{The parameter grid and its cells. When the high dimensional input is fed into the MLP, it will also be divided into images patches of equal size to the cells.}
\label{FIG_GRID}
\end{figure}

Moreover, if the parameter is dynamically modified for each pixel, it will make the MLP more adaptive to image contents. However, predicting a parameter vector for each pixel independently would be computation-intense.
Instead, the hypernetwork $H(\cdot)$ outputs a parameter grid, each cell of which has $S\times$ smaller resolution than the RGB image $X$. Hence, the parameter grid is divided into $S\times S$ cells, so that
\begin{equation}\label{eq10}
\theta = [\theta_{cell_i}]_{i=1, ..., S\times S}.
\end{equation}
Thus, each cell $\theta_{cell}$ of the parameter grid is spatially-varying depending on corresponding patch of the high dimensional input.
In this way, our hypernetwork obtains a very low resolution representation from the multiple samples (e.g., only 16 × 16 pixels for the output with factor $S=4$ as Fig. \ref{FIG_GRID} shows), which dramatically corresponds to image patches of different sizes.
Similar to dynamic filters, it enables our model content-adaptive and our reconstruction performance based on small patches more subtle.

\subsection{Training and implementation details}
The process of our model optimization is minimizing the following loss function:
\begin{equation}\label{eq11}
\theta = \arg \min_{\theta } \sum_{}^{p\in R^{W*H}}|f(X_p, \gamma (p); \theta )-Y(p)|,
\end{equation}
where $X$ and $Y$ denote the pair of RGB image and the ground truth HSI, respectively. $f(\cdot;\theta)$ represents the reconstructed HSI by implicit neural representation with parameters $\theta$. Besides, our experiments show that L1 loss is more adaptive to the optimization than the mean square error loss function. We employ the ADAM optimizer to train the network with initial learning rates $0.0001$.We terminate the training process in 1,000 epochs.

\section{Experimental Results and Discussion}

\subsection{Datasets}

\subsubsection{CAVE}
The CAVE \cite{yasuma2010generalized} dataset is comprised of 32 scenes of a wide range of materials and objects, such as skin, fruits, drinks, feathers, paintings, etc. Each image is collected with 31 spectral bands in a wavelength range of 400-700nm at 10nm interval. The resolutions of hyperspectral images are $512\times 512$. Besides, the CAVE is captured by a tunable filter and a Cooled CCD camera called Apogee Alta U260 under controlled conditions of illumination.

\subsubsection{NUS}
The NUS \cite{nguyen2014training} dataset is captured by a Specim’s PFDCL-65-V10E spectral camera. Specially, different illumination conditions are simulated using metal halide lamps of different color temperatures, commercial LED and natural light sources. There are 66 spectral images and their corresponding illumination spectrum. Similar to the CAVE, the NUS also contains 31 bands between 400 to 700nm at a step of 10nm. The spatial resolutions of all HSIs are beyond $1000 \times 1000$.

\subsubsection{NTIRE2018}
The NTIRE2018 \cite{arad2018ntire} dataset is acquired by a Specim PS Kappa DX4 hyperspectral camera which realizing spatial scanning through a rotary stage. At this time it contains 261 images which are collected at 1392 × 1300 resolution and 31 spectral bands. As yet these data constitutes the largest dataset and it is initially used for a challenge on spectral reconstruction from RGB Images.

\subsection{Evaluation Metrics}
In order to evaluate the performance of HSI super-resolution methods, we utilize three widely used metrics, including Peak Signal to Noise Ratio (PSNR), structural similarity (SSIM), and spectral angle mapping (SAM).
The PSNR is the ratio between the maximum possible power of an image and the power of  distorting noise that affects the quality of its reconstruction:
\begin{equation}\label{eq12}
PSNR(Y, \hat{Y})=10log_{10}(\frac{max(Y)^2}{\frac{1}{W\times H} \left \| Y-\hat{Y}  \right \|_2^2 }).
\end{equation}
It is suggested that the higher the PSNR, the better the matching between the reconstructed HSI and the ground truth image, and the better the reconstruction algorithm. The SSIM is based on visible structures in the image, and can consider image degradation as perceived change in structural information:
\begin{equation}\label{eq13}
SSIM(Y, \hat{Y})=\frac{(2 \mu_{Y} \mu_{\hat{Y}}+c_1)(2\sigma_{Y\hat{Y}}+c_2 )}{(\mu_{Y}^2+\mu_{\hat{Y}}^2+c_1)(\sigma_{Y}^2+\sigma_{\hat{Y}}^2+c_2)},
\end{equation}
where $\mu$ denotes the average value, and $\sigma$ is the variance or covariance, $c_1$ and $c_2$ are two variables to stabilize the result of division operation with weak denominator. Besides, the SAM is a determines the similarity between two spectra by calculating the angle between the estimated HSI and the reference one:
\begin{equation}\label{eq14}
SAM(Y, \hat{Y})=arc cos\frac{<Y^T\hat{Y}>}{\left \| Y \right \|_2\cdot\left \| \hat{Y}  \right \|_2}.
\end{equation}
In general, PSNR, SSIM and SAM are computed on each channel and averaged over all spectral bands. Note that unlike PSNR and SSIM, SAM is inversely proportional to the reconstruction quality.

\begin{figure*}[!t]
  \centering
  \includegraphics[scale=0.7]{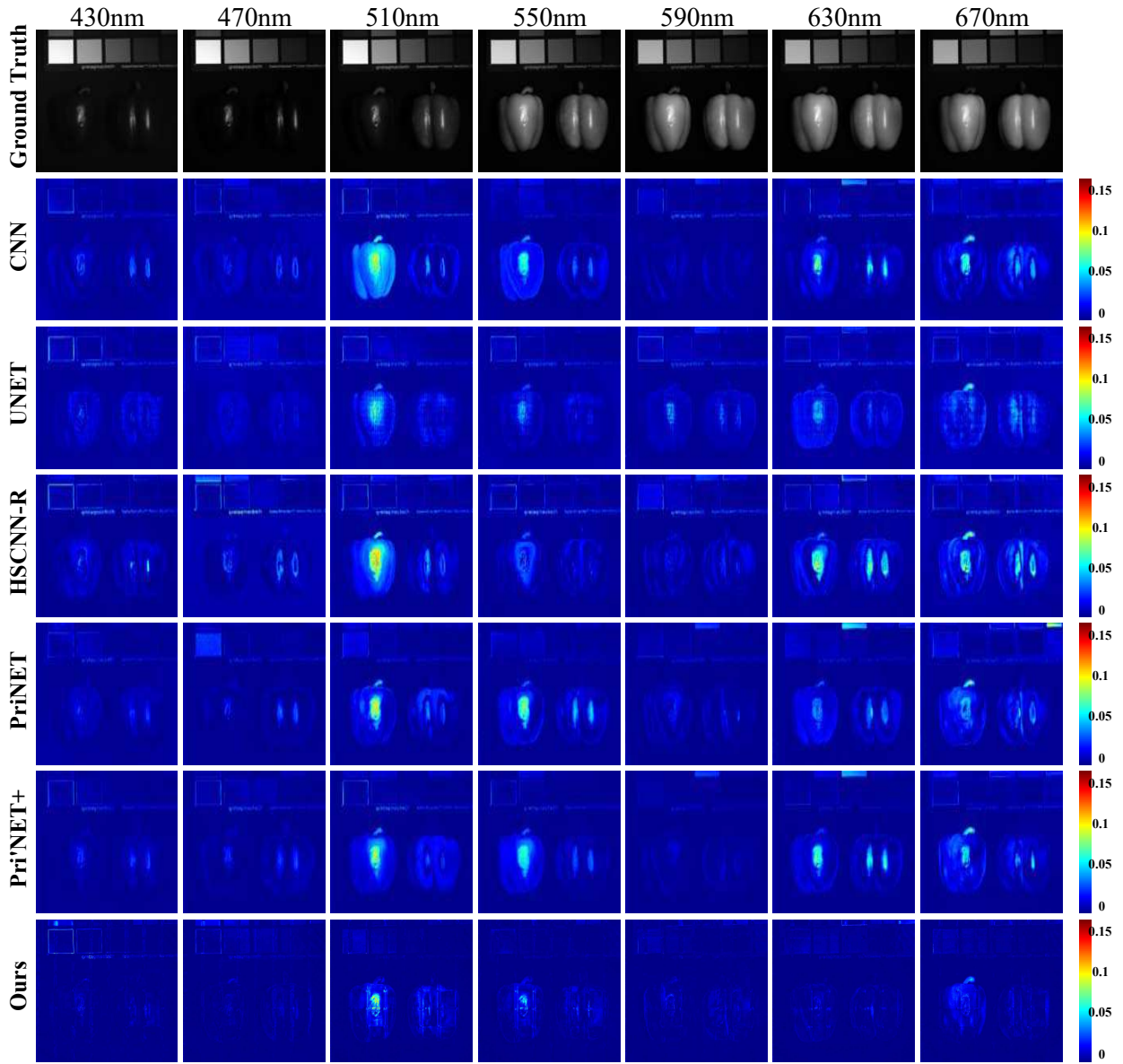}
  \caption{Reconstructed images of "real and fake peppers" on the CAVE dataset at 430, 470, 510, 550, 590, 630, and 670nm. The first row shows the ground truth spectral band, and the other rows show the difference map between the ground truth and the reconstructed result generated by the competing methods and our proposed model.}
\label{FIG_VISUAL_CAVE}
\end{figure*}

\subsection{Implementation Details}
In our experiments, all the images of three datasets are splitted into three parts, among which, 60\% of the images are used for training, 20\% of the images are used to constitute the validation set, and the rest 20\% of the image are for testing. We augment the input images, by randomly sampling a certain amount of $64 \times 64$ patches from each original image. In this way, the number of the input images is expanded to 1000 for the CAVE and NUS dataset and 500 for the NTIRE2018 dataset. We implement our model through PyTorch and all experiments are conducted on a platform with one NVIDIA GeForce RTX 2080Ti GPU and Intel i7-9700K. We adopt Adam optimizer to train our model for 1000 epochs. The initial learning rate is set to $0.0001$ and reduced by a factor 0.1 every 200 epochs.

\subsection{Performance Evaluation}
The proposed model is compared with several state-of-the-art HSI super-resolution methods from single RGB, including CNN based network \cite{fu2018joint}, UNET based network \cite{nie2018deeply}, HSCNN-R based network \cite{shi2018hscnn+}, Multi-scale CNN method \cite{yan2018accurate}, PriNET \cite{hang2020prinet}, and PriNET+ \cite{hang2021spectral}. Since our proposed model is an implement of implicit neural representation, it is abbreviated to INR-based model for simplicity.

\begin{table*}[!ht]
\centering
\caption{Quantitative results of different methods on CAVE, NUS, and NTIRE2018 datasets.}
\begin{tabular}{|c|ccc|ccc|ccc|}
\hline
\multirow{2}{*}{Model} & \multicolumn{3}{c|}{CAVE}                                                                      & \multicolumn{3}{c|}{NUS}                                                                       & \multicolumn{3}{c|}{NTIRE2018}                                                                 \\ \cline{2-10} 
                       & \multicolumn{1}{c|}{PSNR}             & \multicolumn{1}{c|}{SSIM}            & SAM             & \multicolumn{1}{c|}{PSNR}             & \multicolumn{1}{c|}{SSIM}            & SAM             & \multicolumn{1}{c|}{PSNR}             & \multicolumn{1}{c|}{SSIM}            & SAM             \\ \hline
CNN \cite{fu2018joint}                   & \multicolumn{1}{c|}{32.2165}          & \multicolumn{1}{c|}{0.9706}          & 10.7111         & \multicolumn{1}{c|}{25.5296}          & \multicolumn{1}{c|}{0.9238}          & 9.4873          & \multicolumn{1}{c|}{45.8232}          & \multicolumn{1}{c|}{0.9998}          & 1.7232          \\ \hline
UNET \cite{nie2018deeply}                  & \multicolumn{1}{c|}{31.6973}          & \multicolumn{1}{c|}{0.9488}          & 13.1052         & \multicolumn{1}{c|}{25.0038}          & \multicolumn{1}{c|}{0.8872}          & 10.0554         & \multicolumn{1}{c|}{38.2989}          & \multicolumn{1}{c|}{0.9976}          & 3.0454          \\ \hline
HSCNN-R \cite{shi2018hscnn+}                & \multicolumn{1}{c|}{31.4676}          & \multicolumn{1}{c|}{0.9637}          & 12.2081         & \multicolumn{1}{c|}{25.1326}          & \multicolumn{1}{c|}{0.9213}          & 9.5254          & \multicolumn{1}{c|}{45.7062}          & \multicolumn{1}{c|}{0.9998}          & 1.6455          \\ \hline
Multi-scale CNN \cite{yan2018accurate}       & \multicolumn{1}{c|}{31.9298}          & \multicolumn{1}{c|}{0.9575}          & 11.8746         & \multicolumn{1}{c|}{25.1922}          & \multicolumn{1}{c|}{0.9219}          & 9.5021          & \multicolumn{1}{c|}{45.7752}          & \multicolumn{1}{c|}{0.9998}          & 1.6938          \\ \hline
PriNET \cite{hang2020prinet}                 & \multicolumn{1}{c|}{32.8129}          & \multicolumn{1}{c|}{0.9733}          & 10.0400         & \multicolumn{1}{c|}{25.2622}          & \multicolumn{1}{c|}{0.9368}          & 9.9859          & \multicolumn{1}{c|}{46.2661}          & \multicolumn{1}{c|}{0.9999}          & 1.5560          \\ \hline
PriNET+ \cite{hang2021spectral}               & \multicolumn{1}{c|}{32.8300}          & \multicolumn{1}{c|}{\textbf{0.9833}} & 8.7750          & \multicolumn{1}{c|}{26.2893}          & \multicolumn{1}{c|}{0.9405}          & 8.9923          & \multicolumn{1}{c|}{\textbf{46.3500}} & \multicolumn{1}{c|}{0.9999}          & 1.5316          \\ \hline
INR(Ours)              & \multicolumn{1}{c|}{\textbf{34.6257}} & \multicolumn{1}{c|}{0.9781}          & \textbf{7.3285} & \multicolumn{1}{c|}{\textbf{26.3431}} & \multicolumn{1}{c|}{\textbf{0.9568}} & \textbf{8.8187} & \multicolumn{1}{c|}{46.0808}          & \multicolumn{1}{c|}{\textbf{0.9999}} & \textbf{1.5198} \\ \hline
\end{tabular}
\label{Table_results}
\end{table*}

\begin{figure*}[htbp]
  \centering
  \includegraphics[scale=0.53]{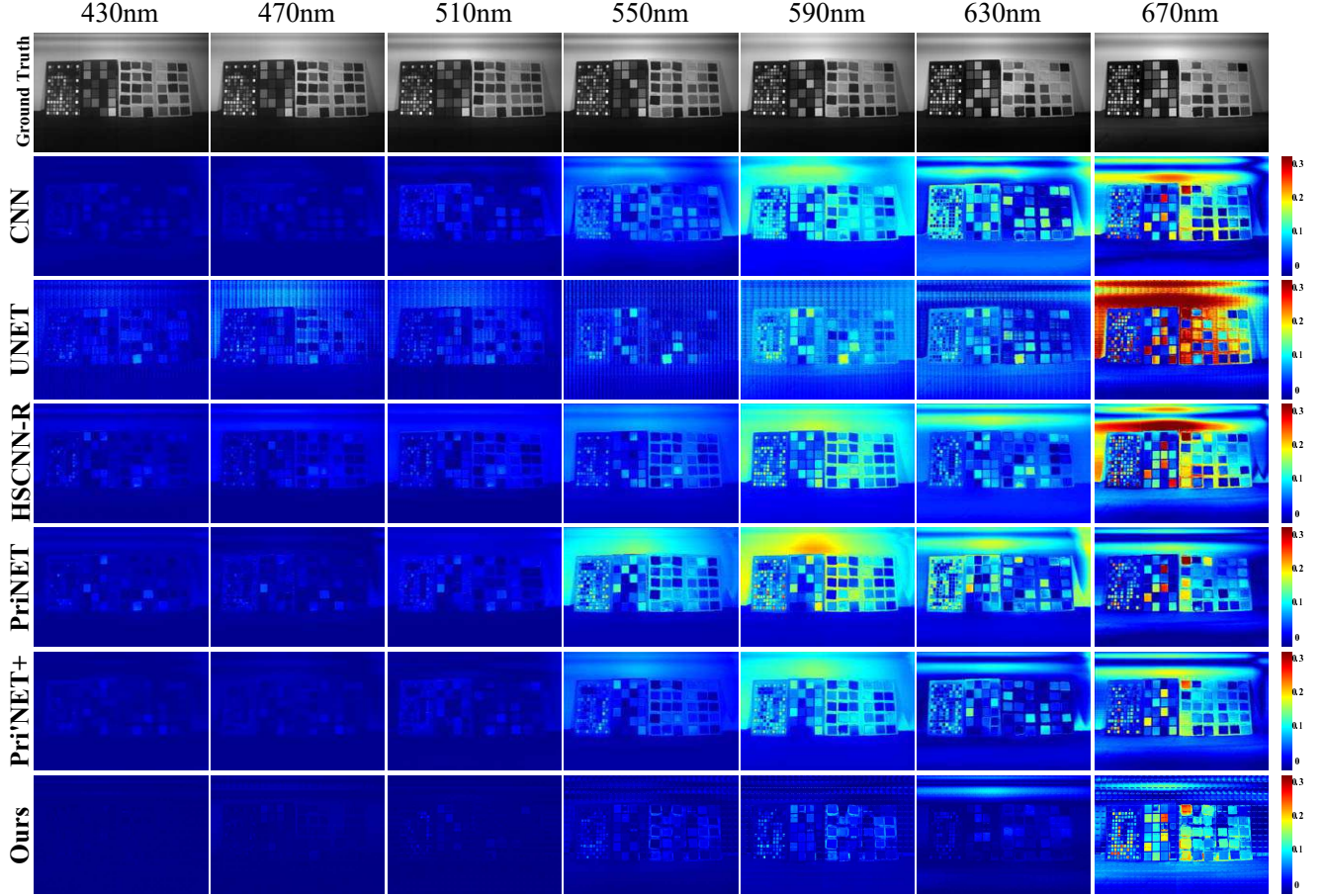}
  \caption{Reconstructed images of "Metal-halide-lamp-2500K-Scene" on the NUS dataset at 430, 470, 510, 550, 590, 630, and 670nm. The first row shows the ground truth HSI, and the other rows show the difference maps between the ground truth and the reconstructed results generated by the competing methods and our proposed model.}
\label{FIG_VISUAL_NUS}
\end{figure*}

\subsubsection{Experimental Results on CAVE Dataset}

The CAVE column of Table \ref{Table_results} summarizes the quantitative reconstruction performances of our method and other methods on the CAVE dataset, and the best results are highlighted in bold font. Our approach results in 34.6257 and 7.3285 in terms of PSNR and SAM, respectively. Compared with other state-of-the-art methods on the CAVE dataset, our method achieves the best performance. More specifically, the PSNR value has a 1.7957 increase and the SAM value is reduced by 1.4465. As for the SSIM value of 0.9781, the performance of the proposed method can be comparable to those of the state-of-the-art methods.

For a more intuitive comparison, we also show some visualization results of "real and fake peppers" from the CAVE dataset. Fig. \ref{FIG_VISUAL_CAVE} illustrates the reconstruction results and the ground truth on seven selected bands with wavelengths of 430, 470, 510, 550, 590, 630, and 670nm. The top row shows the ground truth HSI in each band, while the other rows present the difference maps between the ground truth HSI and reconstruction results generated by six super-resolution methods.
Through contrastive analysis, it can be observed that most of the reconstruction errors are concentrated on the illuminated area, especially for the CNN, HSCNN-R and PriNET+ approaches.
The brighter regions usually present sharp changes in reflections and shadows, which can show textures and lighting effects to reveal more details.
As mentioned, our INR-based method is able to learn sufficient high frequency features, such as sharp edges and fine-grained textures from the input images.
It is so pronounced that the highly reflective areas of 630nm and 670nm bands achieve superior results than other state-of-the-art methods. 
Meanwhile, for the difference maps within other spectrum bands, our INR-based method still has a better reconstruction performance with respect to the compared methods.
Therefore, all of these results validate the effectiveness of our proposed method on the CAVE dataset.

\subsubsection{Experimental Results on NUS Dataset}
The quantitative results of different methods on the NUS dataset are listed in the NUS columns of Table \ref{Table_results}, and the best results are highlighted in bold font.
Our method attains excellent performance which achieves the PSNR of 26.3431, the SSIM of 0.9568, and the SAM of 8.8197.
Compared with those state-of-the-art methods, our method achieves an average of 0.0538 increase in PSNR and an average of 0.0163 increase in SSIM.
Surprisingly, our method gains a tiny advantage over general convolutional networks on the NUS dataset.

Similarly, we also show a qualitative comparison on the NUS dataset in Fig. \ref{FIG_VISUAL_NUS}. The difference maps of selected image named "Metal-halide-lamp-2500K-Scene", including seven bands of 4, 8, 12, 16, 20, 24, and 28, are displayed.
We can observe that the images reconstructed by our method exhibit obvious texture information, especially in color blocks. Moreover, highlight lights reflected on the white background obtain better image restoration effect.
In summary, the reconstruction results can further validate the effectiveness of our proposed method.

\subsubsection{Experimental Results on NTIRE2018 Dataset}
As shown in the NTIRE2018 columns of Table \ref{Table_results}, the quantitative results on the NTIRE2018 dataset are reported.
Our method attains great performance that achieves the PSNR of 46.0808, the SSIM of 0.9999, and the SAM of 1.5198.
More specifically, the SSIM value is approximately equal to the state-of-the-arts, and the SAM value has a 0.0118 increase than PriNET+. As for the PSNR value, the performance of the proposed method is just close to the best result.

\subsection{Ablation Study}
\label{ablation}
In this section, to validate the effectiveness of our proposed method, we conduct extensive ablation experiments on the CAVE dataset. The effect of different factors on parametric model of INR by comparing the reconstruction performances are clarified.

\begin{table}[!b]
\caption{The reconstruction results on different sizes of parameter grid.}
\centering
\begin{tabular}{cccccc}
\hline
S    & \#Ds layers & Params & PSNR             & SSIM            & SAM             \\ \hline
S=2  & 5           & 64.9M  & \textbf{33.8938} & 0.9718          & 7.5979          \\
S=4  & 4           & 64.9M  & 33.1491          & 0.9729          & 7.5561          \\
S=8  & 3           & 64.8M  & 33.4025          & \textbf{0.9730} & 7.6563          \\
S=16 & 2           & 64.8M  & 33.8625          & 0.9702          & \textbf{7.5359} \\ \hline
\end{tabular}
\label{Table_ab1}
\end{table}

\begin{figure}[!htbp]
  \centering
  \includegraphics[scale=0.45]{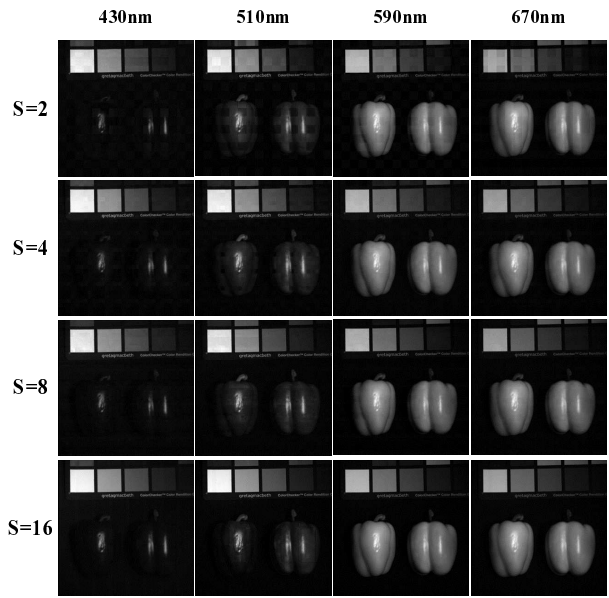}
\caption{Reconstructed images of "real and fake peppers" generated by our INR model with different $S$ values on the CAVE dataset at 430nm, 510nm, 590nm, and 670nm spectral bands.}
\label{FIG_AB1}
\end{figure}

\begin{table}[!b]
\caption{Reconstruction results on different frequencies of periodic spatial encoding.}
\centering
\begin{tabular}{ccccc}
\hline
Per. Spa. Enc. & S    & PSNR             & SSIM            & SAM             \\ \hline
w/o            & S=16 & 32.6752          & 0.9669          & 8.3312          \\
N=3            & S=16 & 34.0027          & 0.9747          & 7.7469          \\
N=5            & S=16 & \textbf{34.6257} & \textbf{0.9781} & \textbf{7.3285} \\ \hline
\end{tabular}
\label{Table_ab2}
\end{table}

\subsubsection{Discussion on Different Sizes of Parameter Grid}

As discussed in Section \ref{section_grid}, the weights and bias of MLP are divided into a parameter grid by the scaling factor $S$. $S$ determines the size of the output of the hypernetwork, and actually, the number of downsampling layers in hypernetwork is directly relative to $S$. Accordingly, the MLP processes a low resolution patch of the high dimensional input (e.g., only 16 × 16 pixels for image patches with the factor $S=4$).

Table \ref{Table_ab1} summarizes the reconstruction performances of our method with different size of parameter grid. The best results are highlighted in bold font and it can be observed that the results are similar. However, from the reconstructed images with different $S$ values in Fig. \ref{FIG_AB1}, we can tell the impact of different grid size on image reconstruction effectiveness.
When $S$ is given a smaller value, the image patch and grid size are too large, which often results in heavy blocking artifacts near the grid boundaries. The most obvious is in the 430nm band of Fig. \ref{FIG_AB1} when $S$ is set to 2 and 4. As demonstrated, with the scaling factor $S$ increasing, 
blocking artifacts can be almost eliminated and the reconstruction performance is also improved.
Overall, the finer parameter grid makes the generated images smoother and more realistic.

\subsubsection{Discussion on Different Frequencies of Periodic Spatial Encoding}

As discussed in Section \ref{section_encod}, we perform periodic spatial encoding to map the positional coordinates to a higher dimensional space for enabling the network to learn sufficient high frequency image variation. Thus, we investigate the impact of frequency of periodic spatial encoding.
Starting from omitting the position encoding, the positional coordinates are nomalized to [0,1] and fed as the input without periodic spatial encoding.
Afterwards, we gradually activate the encodings of higher frequency bands with $N=3$ and $N=5$.
The final results are shown in Table \ref{Table_ab2}. We notice that our performance stably increase with the frequency of encoding. When the spatial encoding is not consumed, the corresponding results are far less than others. 
In summary, the periodic spatial encoding with high frequency can largely improve the performance of the generated image.

\section{Conclusion}
In this paper, we propose a novel reconstruction model based on implicit neural representation to improve the performance of hyperspectral image super-resolution task.
The model is to learn the continuous representation function that precisely predicts spectral radiance intensities from input spatial coordinates using a fully connected network.
In particular, the parameters of the fully connected network are generated by a hypernetwork, which makes each image be represented by an individual function, and it can more realistically reflect the actual imaging process.
Moreover, the sinusoidal encodings of positional coordinates are concatenated with the input image to make the model capable of recovering more high-frequency details.
The reconstruction results on CAVE, NUS, and NTIRE2018 datasets indicate the superiority of our proposed model.

\ifCLASSOPTIONcaptionsoff
  \newpage
\fi



%

\bibliographystyle{IEEEtran}
\bibliography{mybib}

%





\end{document}